\documentclass[prl,twocolumn,amsmath,amssymb,superscriptaddress]{revtex4}

\usepackage{graphicx}
\usepackage{dcolumn}
\usepackage{bm}
\usepackage{color}

\begin{document}

\preprint{APS/123-QED}

\title{Laser-Induced Charge-Density-Wave Transient Depinning in Chromium}

\author{V.L.R. Jacques}
\affiliation{Laboratoire de Physique des Solides, CNRS, Univ. Paris-Sud, Universit\'e Paris-Saclay, 91405 Orsay Cedex, France}
 \email{vincent.jacques@u-psud.fr}
\author{C. Laulh\'e}
 \affiliation{Synchrotron Soleil, L’Orme des Merisiers, Saint-Aubin, BP 48, FR-91192 Gif-sur-Yvette Cedex, France}
 \affiliation{Universit\'e Paris-Saclay (PSud), F-91405 Orsay Cedex, France}
\author{N. Moisan}
 \affiliation{Laboratoire de Physique des Solides, CNRS, Univ. Paris-Sud, Universit\'e Paris-Saclay, 91405 Orsay Cedex, France}
\author{S. Ravy}
 \affiliation{Laboratoire de Physique des Solides, CNRS, Univ. Paris-Sud, Universit\'e Paris-Saclay, 91405 Orsay Cedex, France}
\author{D. Le Bolloc'h}
 \affiliation{Laboratoire de Physique des Solides, CNRS, Univ. Paris-Sud, Universit\'e Paris-Saclay, 91405 Orsay Cedex, France}

\date{\today}

\begin{abstract}
We report here on time-resolved x-ray diffraction measurements following femtosecond laser excitation in pure bulk chromium. Comparing the evolution of incommensurate charge-density-wave (CDW) and atomic lattice reflections, we show that, few nanoseconds after laser excitation, the CDW undergoes different structural changes than the atomic lattice. We give evidence for a transient CDW shear strain that breaks the lattice point symmetry. This strain is characteristic of sliding CDWs, as observed in other incommensurate CDW systems, suggesting the laser-induced CDW sliding capability in 3D systems. This first evidence opens perspectives for unconventional laser-assisted transport of correlated charges.
\end{abstract}

\maketitle

Understanding the interplay between spin, charge and lattice is a major issue in condensed matter. Chromium is a typical system having complex electronic and magnetic ground states despite a basic crystallographic structure~\cite{fawcett1988}. Below 311 K, a spin-density-wave appears with twice the period of a charge-density-wave (CDW)/strain wave modulation. In bulk chromium, the ratio of the atomic lattice and CDW periods is incommensurate. In principle, this implies that energetically equivalent states are found whatever the position of the CDW with respect to the atomic lattice. However, this translational invariance, inducing transport of correlated charges in low-dimensional systems, has never been observed in 3D systems like chromium~\cite{parker1991}.

Systems submitted to an external driving force in disordered media share universal behaviours. For various systems, such as surfaces, vortices in type-II superconductors~\cite{Giamarchi1995}, or magnetic domain walls~\cite{Thiaville2005}, similar regimes are sequentially observed - pinning, creep and flow - depending on the pinning strength compared to the applied force magnitude. The case of periodic systems, like CDWs, is peculiar. They are generally found in low dimensional materials, characterized by strong structural anisotropy, when a periodic lattice distortion allows a major decrease of the electron energy thanks to a gap opening, resulting in a static modulation of the electron density~\cite{monceau2012}. CDWs are pinned to the lattice either because of local impurity potentials or commensurability effects between the lattice and the CDW periodicities. Depinning thus requires the CDW and lattice periods to be incommensurate, i.e. to have an irrational ratio. When it takes place, the collective transport of charges is detectable through the non-ohmic behaviour of the current-voltage characteristics, as well as an additional ac voltage. This effect has been observed in several quasi-one dimensional systems like in NbSe3~\cite{Monceau1976} and  in blue bronze~\cite{Dumas1983}. More recently, quasi-2D CDW systems were also found to have this ability~\cite{sinchenko2012,LeBolloch2016} but CDW sliding has never been observed in 3D materials so far.

The presence of a CDW in an isotropic 3D metal like chromium is exceptional. In the bulk, this metal indeed displays incommensurate DWs although its structure is cubic and monoatomic, with hardly any anisotropy of its properties~\cite{fawcett1988}. It was the first metal identified to display a linear incommensurate spin density wave (SDW) with wavevector $\delta$ due to itinerant 3d electrons~\cite{overhauser1962}. Associated charge harmonics were first predicted~\cite{Young1974} and later evidenced~\cite{Tsunoda1974}. The coexistence of both CDW and SDW in chromium has led to extensive studies to unravel the coupling between spin, charge and lattice~\cite{overhauser1962,Corliss1959,Werner1967,Young1974,Tsunoda1974,Cowan1978,Jacques2009b,Jacques2014,Evans2002}.

A temporal study of these components can provide valuable information about their interdependency. Contrary to many other CDW systems~\cite{Yusupov2010,Demsar1999,Tomeljak2009,Schmitt2008,Liu2013,Eichberger2010,huber2014,laulhe2015}, the ultrafast dynamics of chromium has never been studied in bulks. Femtosecond reflectivity experiments have been reported in films, validating that the ultrafast electronic response is well accounted for by the two-temperature model~\cite{hirori2003,Brorson1990}, and confirmed by a recent study of a commensurate CDW in a Cr film as well~\cite{singer2015}.

\begin{figure}[!ht]%
$$\includegraphics[width=\columnwidth]{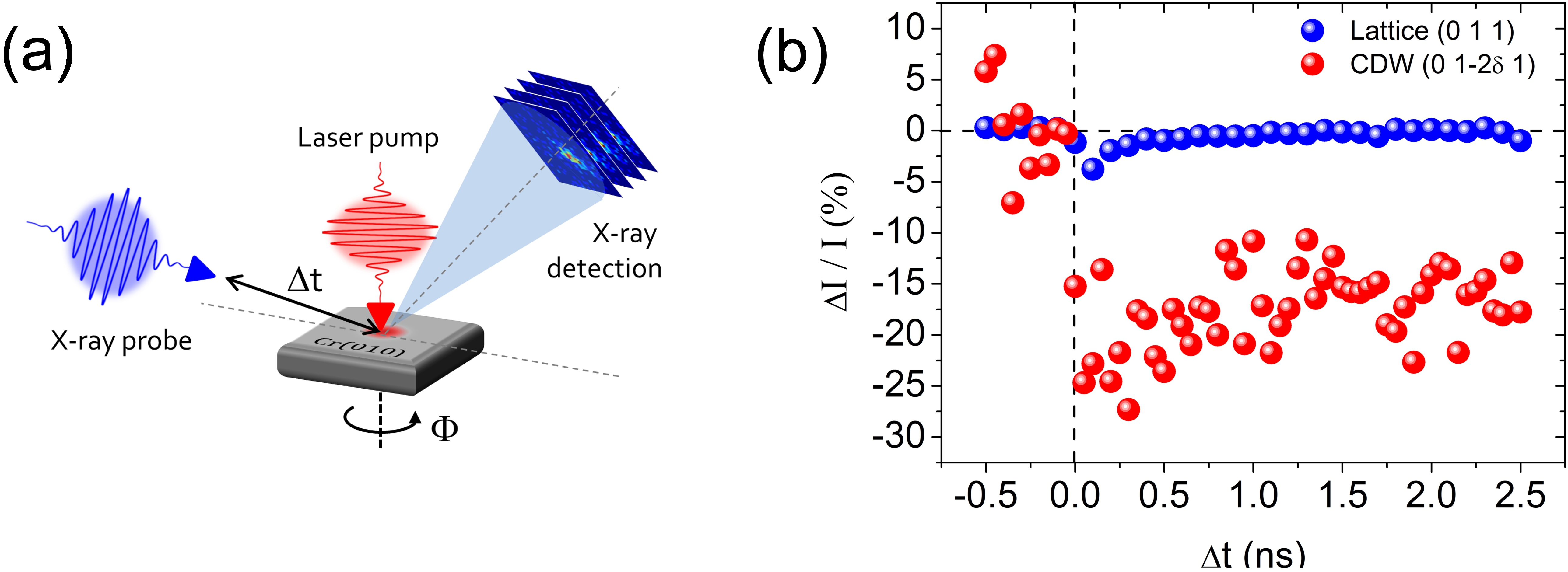}$$%
\caption{(color online) (a) Experimental setup scheme. The Cr(010) single crystal is excited by a 800nm femtosecond laser pulse and probed by a 70 ps x-ray pulse sent with a temporal delay $\Delta$t after laser excitation, in 1$^o$ grazing-incidence geometry. (b) Intensity variations of the (0,1,1) Bragg reflection and of the (0,1-2$\delta$,1) satellite reflection associated to the lattice and the CDW respectively, recorded at fixed position of sample and detector during the first nanoseconds following laser excitation. }
\label{fig1}
\end{figure}

In this work, the CDW modulation (and the superimposed periodic lattice distortion) and the average lattice have been studied as a function of time after femtosecond laser excitation by picosecond time-resolved x-ray diffraction in a bulk. The strength of this technique is its wavevector selectivity, which allows to track the time-dependent behaviour of the CDW and average lattice independently. The experiment was performed at the CRISTAL beamline of the SOLEIL synchrotron, in the 8-bunch operation mode. The setup used for the experiment is shown in Fig.~\ref{fig1}(a). 

The temporal resolution of this experiment was given by the x-ray pulse duration of 76 ps FWHM. A multi-Q (010) Cr single crystal was mounted on the 6-circle diffractometer of the beamline, and excited with 30 fs infrared laser pulses (800 nm wavelength) synchronized with the X-ray pulses at a repetition rate of ~ 1 kHz. The laser beam impinged the sample perpendicular to its (010) surface with a spot size radius of ~2 mm, yielding a fluence of 6.2 mJ/cm$^2$. The (0,1-2$\delta$,1) reflection associated to the CDW was probed with 7.15 keV x-ray pulses (wavelength 1.734 \AA), far from the chromium K-edge energy (E$_K$=5.988 keV)). The (0,1,1) reflection was measured to track variations of the lattice structure. X-rays were set in order to impinge the sample at a grazing angle $\theta_i$ = 1$^o$ - above the 0.414$^o$ critical angle of chromium at this energy - to get an effective penetration depth of 70 nm along the sample surface normal, matching the laser penetration depth coming along the sample normal (see Fig.~\ref{fig1}(a)). The x-ray beam footprint was $\sim$ 1.7 mm along the incident beam direction and 0.5 mm in the perpendicular direction. Detection was performed using a 2D pixel detector (XPAD3.2), located at 387 mm from the sample, leading to a resolution of 1.217 10$^{-3}$ \AA$^{-1}$ in reciprocal space. Single bunches were selected by synchronizing a 100 ns detector counting gate with the laser excitation pulse. Each pump-probe measurement was obtained by summing 1000 events. The sample was cooled down using a He blower refrigerator to reach temperatures from 50 K to 300 K without screening the laser and x-ray beams on the sample. When the laser is switched on, the CDW reflection is found at a different position than the one measured without laser in reciprocal space, and corresponds to an overall temperature increase of around 30K. 

\begin{figure}[!ht]%
\includegraphics[width=\columnwidth]{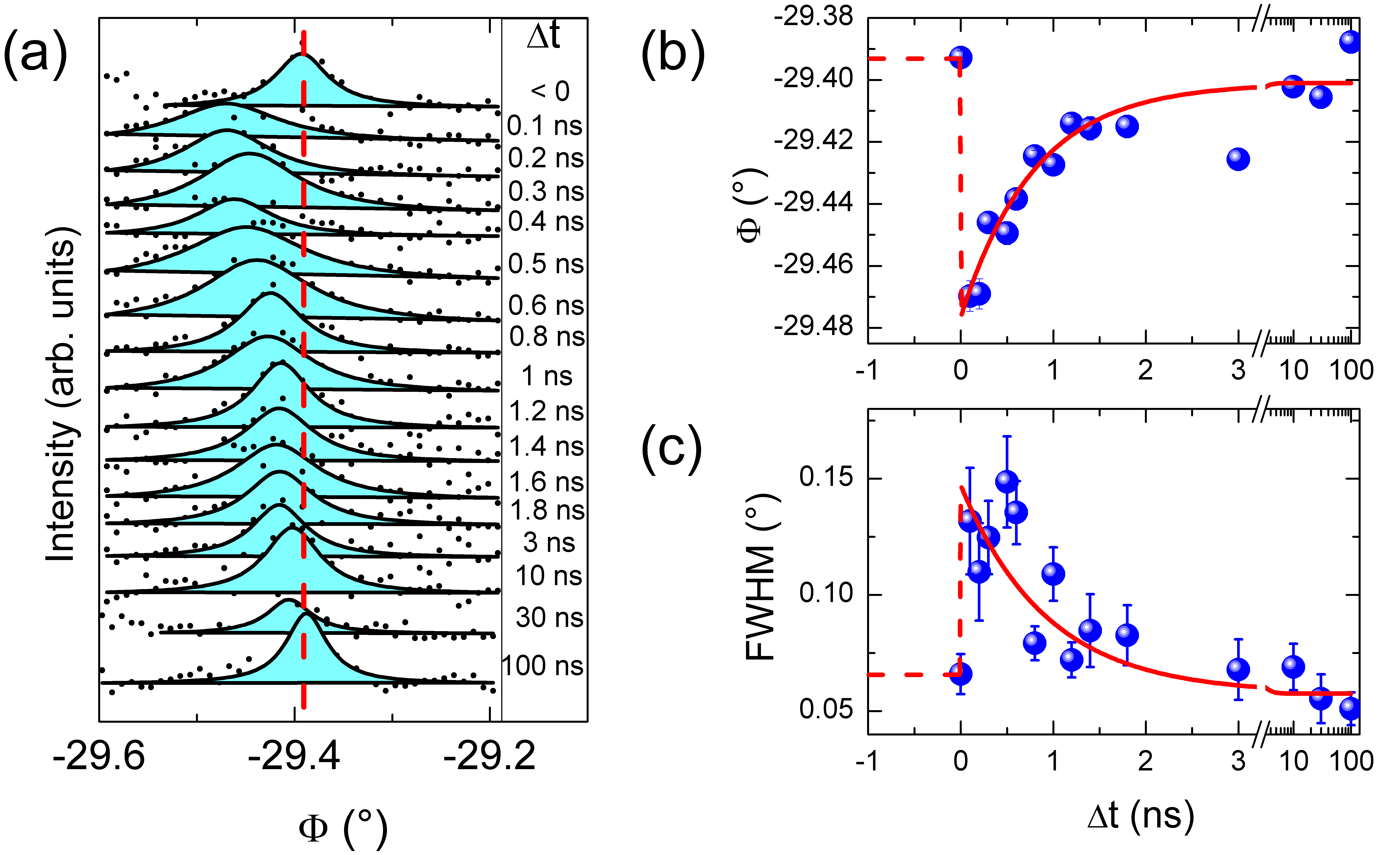}%
\caption{(color online) (a) $\Phi$-scans as a function of time delay $\Delta$t, obtained by integrating the signal on the detector at 140 K. The dots are measured points and the blue filled-areas bound by solid lines are fit of the experimental data. The position of the maximum measured at negative delays is indicated by the red dashed line. (b) Evolution of the central position and (c), of the FWHM of the CDW reflection extracted from the fits shown in a as a function of time delay $\Delta$t (blue dots). The red solid line is exponential fit to the data, and the red dashed lines are guides to the eye.}
\label{fig2}
\end{figure}

A time-resolved measurement was performed at 140 K on the (0,1-2$\delta$,1) satellite reflection associated to the CDW, and on the (0,1,1) fundamental Bragg reflection associated to the underlying cubic crystal structure (see Fig.~\ref{fig1}(b)). While intensity variations of the (0,1,1) Bragg reflection keep smaller than 5\%, a 30\% intensity drop is observed for the CDW satellite peak during the first 250 ps following laser excitation. The initial state is not fully recovered after $\Delta$t = 2.5 ns.
In order to get full information on the CDW satellite peak position in reciprocal space, we performed azimuthal angle $\Phi$-scans for different pump-probe delays (see Fig.~\ref{fig1}(a)). The time-evolutions of the CDW peak position and Full Width at Half Maximum (FWHM) are plotted in Fig.~\ref{fig2}(b) and (c) respectively.

At $\Delta$t = 0.1 ns after laser excitation, the CDW peak position is shifted with respect to its negative delay value, and accompanied by a clear peak broadening. The diffraction angle and FWHM then continuously relax until reaching their initial value at $\Delta$t $\sim$ 100 ns. Diffraction angle and FWHM follow similar exponential evolution with a characteristic time of ~3 ns to recover 90\% of the initial values. Maximum changes are found at $\Delta$t = 0.1 ns with a peak shifted by 0.08$^o$ and twice as broad as before excitation. The corresponding correlation lengths are 0.4 $\mu$m in the initial state and 0.2 $\mu$m at $\Delta$t = 0.1 ns. In comparison the (0,1,1) Bragg peak position and shape hardly change (see Supplemental Material).
This measurement first shows that the CDW is excited by the laser pulse in a time shorter than the 70 ps time resolution of this experiment and is then followed by a much longer relaxation process involving a strong change of CDW correlation lengths.

\begin{figure}[!ht]%
\includegraphics[width=\columnwidth]{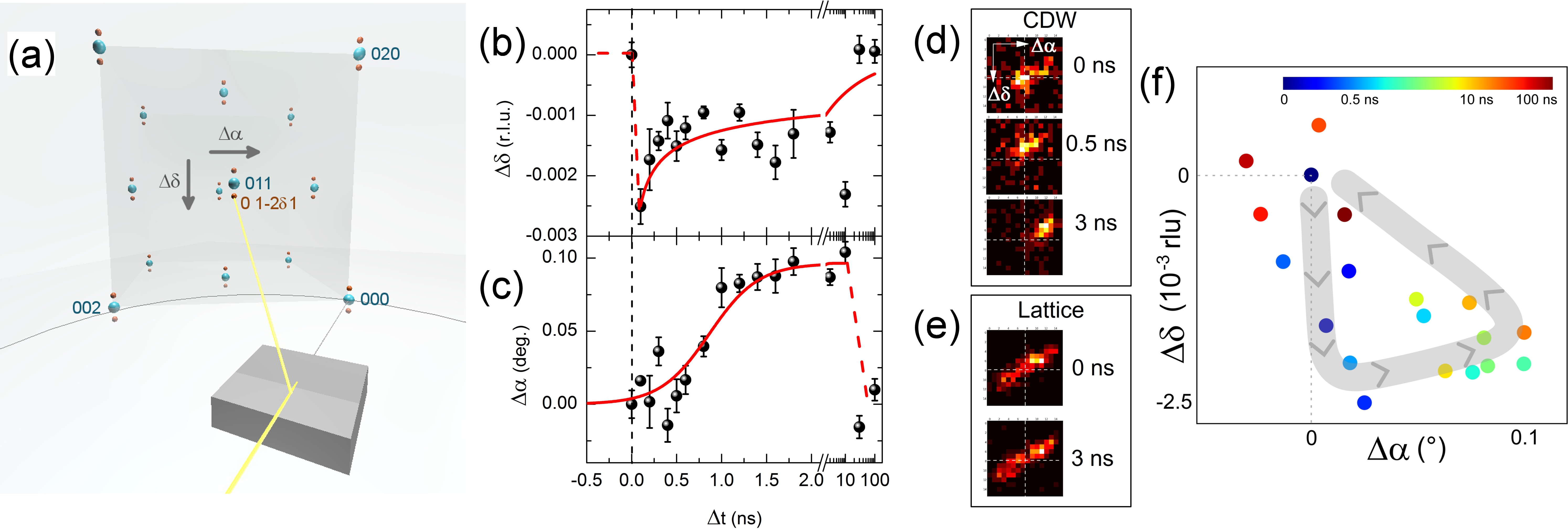}%
\caption{(color online) (a) Sketch of the Ewald construction to scale in reciprocal space (see Supplemental Material), showing the incident beam impinging the sample with a grazing angle and the reciprocal space geometry when the (0,1-2$\delta$,1) reflection is in diffraction condition. Blue dots are reciprocal space points associated to the lattice, and orange ones to the CDW. CDW wavevector variations $\Delta\delta$ appear along the vertical direction of recorded images, and tilts $\Delta\alpha$ along the horizontal direction. (b) Variations of the CDW wavevector $\delta$ and (c) of its angle $\alpha$ relative to their values at negative delay, extracted from the peak position on the detector taken on the maximum of the $\Phi$-scans shown in Fig. 2(a). Black dots are experimental data, red lines fits to the data. (d) CDW reflection recorded on the 2D detector at $\Delta$t = 0 ns; 0.5 ns and 3 ns at the maximum $\Phi$. (e) (0,1,1) lattice reflection recorded at $\Delta$t = 0 ns and 3 ns at the same position of the sample as the CDW shown in (d). No position variation is observed for the lattice reflection. (f) Evolution of the CDW reflection in reciprocal space for delays between 0 and 100 ns. Each dot represents the CDW reflection center of mass as recorded on the 2D detector, and the color code indicating the delay is shown on the colorscale on the top of the image. The grey line depicts the global trajectory of the CDW pic on the detector, and arrows its direction.}
\label{fig3}
\end{figure}

The most surprising point is that this relaxation process also involves a deep modification of the CDW structure. The detailed q-analysis reveals a peculiar behaviour of the CDW wavevector during this relaxation, involving a dilatation-contraction as well as a tilt. This is clearly observed by vertical shifts $\Delta\delta$ on the detector corresponding to variations of the wavevector longitudinal component $\delta$ and horizontal shifts $\Delta\alpha$ associated to tilts of the CDW wavevector (see Fig.~\ref{fig3}(a)). The temporal evolution of $\Delta\delta$ and $\Delta\alpha$ extracted from the images recorded at the maximum intensity of the $\Phi$-scan for each delay is plotted in Fig.~\ref{fig3}(b)-(c). 
\onecolumngrid

\twocolumngrid

The temporal evolution follows three successive steps: from 0 to 500 ps, the longitudinal component $\Delta\delta$ abruptly changes, from 500 ps to 10 ns a strong $\Delta\alpha$ variation induces a lattice point symmetry breaking, and relaxation occurs after 10 ns. During the first step, the longitudinal component of the CDW wavevector $\delta$ decreases in less than 70 ps, and corresponds to a CDW expansion in quantitative agreement with the $\Phi$ time-dependence shown in Fig.~\ref{fig2}(b). In the relaxed state, $\delta$ $\sim$ 0.0447 reciprocal lattice units (r.l.u.), whereas at $\Delta$t = 0.1 ns $\delta$ = 0.0422 r.l.u. This corresponds to an effective temperature increase of 30 K due to laser excitation according to thermodynamical measurements~\cite{Werner1967}. The CDW in the excited part of the sample expands faster than 70 ps while in the non-excited regions it keeps the same period, with a nearly 6\% mismatch between the two regions. The huge associated CDW strain must lead to the nucleation of necessary dislocations in the boundary region and glide during the out-of-equilibrium process to accommodate for the strain variation. This mechanism could be responsible for the decrease of the CDW correlation length at the nanosecond timescale. 

However the evolution from 500 ps to 10 ns clearly shows that the return to equilibrium does not simply involve the reversible contraction of the CDW and a recovery of initial correlation lengths, but that an intermediate step takes place. In this temporal window, the CDW wavevector continuously rotates up to 0.08$^o$ from 0.5 to 10 ns, while no change of the fundamental Bragg refection is observed (Fig.~\ref{fig3}(e)). 

This substantial bending is not expected in the CDW relaxation process as it breaks the orientational symmetry of the CDW with respect to the lattice: the CDW wavevector is no more collinear with the [010] direction of the cubic lattice. This result has fundamental implication since, despite the CDW incommensurability that implies translational invariance of the system, no CDW depinning has ever been observed in chromium~\cite{parker1991}.

All possible experimental asymmetries prove inconsistent with the observed phenomenon. A first possible asymmetry in the experiment is the laser incidence on the sample, which could deviate slightly for the sample normal. The incidence angle incertitude is ~1$^o$ around the normal incidence. With a 2 mm diameter spot, this corresponds to a 35 $\mu$m propagation difference, and thus to 100 fs time difference, between excited regions on both sides of the laser spot, incompatible with the timescales involved here. Phonons cannot be invoked either, as their propagation at $\sim$5.10$^3$ m/s within the 100 nm probed region takes place within 20 ps, much faster than the ns timescale at which the CDW tilt occurs. 

\begin{figure}[!ht]%
\includegraphics[width=\columnwidth]{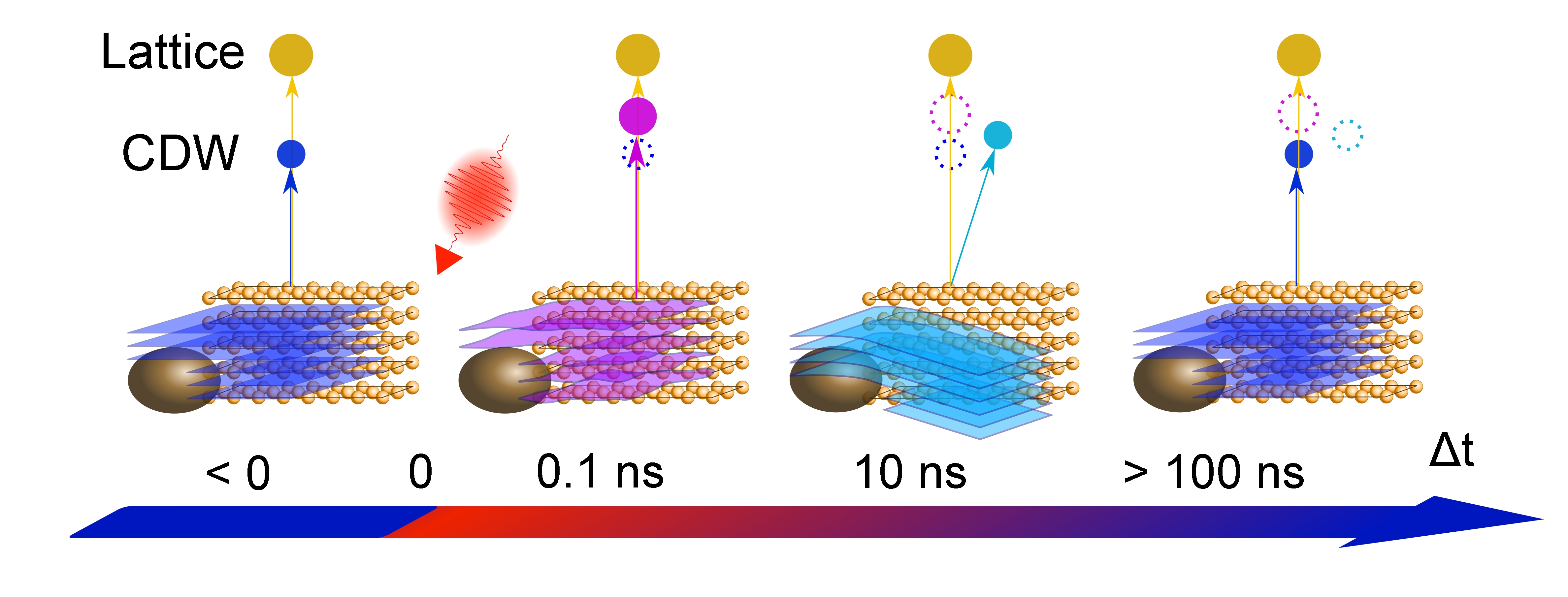}%
\caption{(color online) Time evolution of the system after laser excitation near a pinning center. The bottom time line specifies the relative time-delay between x-ray and laser pulses. Blue portions depict the equilibrium state of the system and red ones to excited warmer states. The probed sample portion is depicted in the middle line, and is not to scale. The yellow balls grid describes the atomic lattice and the probed CDW domains oriented along the surface normal are the parallel sheets (not to scale), blue in the thermodynamical state ($\Delta$t $<$ 0 and $\Delta$t $>$ 100ns), purple at $\Delta$t = 0.1 ns and light blue at $\Delta$t = 10 ns. The brown region depicts a pinning center. In the upper part of the figure, the time-evolution of the lattice and CDW reflection are drawn and follow the orientation and period changes of the structures in real space and at each delay. The open circles in dashed colour lines remind the peak positions at earlier delays. At negative delays, the CDW has long correlation lengths, and is oriented along the surface normal. At $\Delta$t = 1 ns, the CDW period is larger, and its wavefronts are distorted, leading to shorter correlation lengths. At $\Delta$t = 10 ns, the wavefronts are less distorted, the CDW period is changing towards its equilibrium period, but pinning on domains leads to a rotation of its wavevector, and is no longer collinear with the atomic lattice orientation, which breaks the system symmetry. The system is back to its original state after 100 ns.}
\label{fig4}
\end{figure}

The phenomenon involved in the observed symmetry breaking is linked to a shear of the CDW as predicted theoretically~\cite{Feinberg1988} and observed in sliding CDWs~\cite{Isakovic2006}. A possible scenario relies on the interaction of the CDW with a strong pinning center localized close to the probed volume as shown in Fig.~\ref{fig4}. During the out-of-equilibrium process, the CDW wavelength increases while it is pinned on one side, making it bend. Note that the diffraction patterns shown in Fig.~\ref{fig3} imply a global CDW rotation probably due to incompressibility of the CDW due to Coulomb repulsion.

The tilt of the CDW wavevector while the lattice keeps constant indicates that the CDW undergoes a shear strain. Shear is a feature shared by many sliding CDW systems. Observing shear in the out-of-equilibrium CDW and not on the atomic lattice shows that the CDW can depin from the atomic lattice, similar to other incommensurate CDWs found in low-dimensional systems~\cite{Monceau1976,Dumas1983,sinchenko2012}. Furthermore, depinning is induced by an ultrashort laser excitation here, instead of a dc current for usual sliding CDW systems. Our results provide new perspectives in terms of correlated charge transport using laser excitations to induce CDW sliding. 

The authors would like to acknowledge Soleil Synchrotron for providing beamtime and for support during experiment. 

\bibliographystyle{apsrev-nourl-noissn}
\bibliography{cr_CDW_dynamics}

\begin{thebibliography}{31}
\expandafter\ifx\csname natexlab\endcsname\relax\def\natexlab#1{#1}\fi
\expandafter\ifx\csname bibnamefont\endcsname\relax
  \def\bibnamefont#1{#1}\fi
\expandafter\ifx\csname bibfnamefont\endcsname\relax
  \def\bibfnamefont#1{#1}\fi
\expandafter\ifx\csname citenamefont\endcsname\relax
  \def\citenamefont#1{#1}\fi
\expandafter\ifx\csname url\endcsname\relax
  \def\url#1{\texttt{#1}}\fi
\expandafter\ifx\csname urlprefix\endcsname\relax\def\urlprefix{URL }\fi
\providecommand{\bibinfo}[2]{#2}
\providecommand{\eprint}[2][]{\url{#2}}

\bibitem[{\citenamefont{Fawcett}(1988)}]{fawcett1988}
\bibinfo{author}{\bibfnamefont{E.}~\bibnamefont{Fawcett}},
  \bibinfo{journal}{Rev. Mod. Phys.} \textbf{\bibinfo{volume}{60}},
  \bibinfo{pages}{209} (\bibinfo{year}{1988}).

\bibitem[{\citenamefont{Parker and Zettl}(1991)}]{parker1991}
\bibinfo{author}{\bibfnamefont{I.~D.} \bibnamefont{Parker}} \bibnamefont{and}
  \bibinfo{author}{\bibfnamefont{A.}~\bibnamefont{Zettl}},
  \bibinfo{journal}{Phys. Rev. B} \textbf{\bibinfo{volume}{44}},
  \bibinfo{pages}{5313} (\bibinfo{year}{1991}).

\bibitem[{\citenamefont{Giamarchi and Le~Doussal}(1995)}]{Giamarchi1995}
\bibinfo{author}{\bibfnamefont{T.}~\bibnamefont{Giamarchi}} \bibnamefont{and}
  \bibinfo{author}{\bibfnamefont{P.}~\bibnamefont{Le~Doussal}},
  \bibinfo{journal}{Phys. Rev. B} \textbf{\bibinfo{volume}{52}},
  \bibinfo{pages}{1242} (\bibinfo{year}{1995}).

\bibitem[{\citenamefont{Thiaville et~al.}(2005)\citenamefont{Thiaville,
  Nakatani, Miltat, and Suzuki}}]{Thiaville2005}
\bibinfo{author}{\bibfnamefont{A.}~\bibnamefont{Thiaville}},
  \bibinfo{author}{\bibfnamefont{Y.}~\bibnamefont{Nakatani}},
  \bibinfo{author}{\bibfnamefont{J.}~\bibnamefont{Miltat}}, \bibnamefont{and}
  \bibinfo{author}{\bibfnamefont{Y.}~\bibnamefont{Suzuki}},
  \bibinfo{journal}{EPL (Europhysics Letters)} \textbf{\bibinfo{volume}{69}},
  \bibinfo{pages}{990} (\bibinfo{year}{2005}).

\bibitem[{\citenamefont{Monceau}(2012)}]{monceau2012}
\bibinfo{author}{\bibfnamefont{P.}~\bibnamefont{Monceau}},
  \bibinfo{journal}{Advances in Physics} \textbf{\bibinfo{volume}{61}},
  \bibinfo{pages}{325} (\bibinfo{year}{2012}).

\bibitem[{\citenamefont{Monceau et~al.}(1976)\citenamefont{Monceau, Ong,
  Portis, Meerschaut, and Rouxel}}]{Monceau1976}
\bibinfo{author}{\bibfnamefont{P.}~\bibnamefont{Monceau}},
  \bibinfo{author}{\bibfnamefont{N.~P.} \bibnamefont{Ong}},
  \bibinfo{author}{\bibfnamefont{A.~M.} \bibnamefont{Portis}},
  \bibinfo{author}{\bibfnamefont{A.}~\bibnamefont{Meerschaut}},
  \bibnamefont{and} \bibinfo{author}{\bibfnamefont{J.}~\bibnamefont{Rouxel}},
  \bibinfo{journal}{Phys. Rev. Lett.} \textbf{\bibinfo{volume}{37}},
  \bibinfo{pages}{602} (\bibinfo{year}{1976}).

\bibitem[{\citenamefont{Dumas et~al.}(1983)\citenamefont{Dumas, Schlenker,
  Marcus, and Buder}}]{Dumas1983}
\bibinfo{author}{\bibfnamefont{J.}~\bibnamefont{Dumas}},
  \bibinfo{author}{\bibfnamefont{C.}~\bibnamefont{Schlenker}},
  \bibinfo{author}{\bibfnamefont{J.}~\bibnamefont{Marcus}}, \bibnamefont{and}
  \bibinfo{author}{\bibfnamefont{R.}~\bibnamefont{Buder}},
  \bibinfo{journal}{Phys. Rev. Lett.} \textbf{\bibinfo{volume}{50}},
  \bibinfo{pages}{757} (\bibinfo{year}{1983}).

\bibitem[{\citenamefont{Sinchenko et~al.}(2012)\citenamefont{Sinchenko, Lejay,
  and Monceau}}]{sinchenko2012}
\bibinfo{author}{\bibfnamefont{A.~A.} \bibnamefont{Sinchenko}},
  \bibinfo{author}{\bibfnamefont{P.}~\bibnamefont{Lejay}}, \bibnamefont{and}
  \bibinfo{author}{\bibfnamefont{P.}~\bibnamefont{Monceau}},
  \bibinfo{journal}{Phys. Rev. B} \textbf{\bibinfo{volume}{85}},
  \bibinfo{pages}{241104} (\bibinfo{year}{2012}).

\bibitem[{\citenamefont{Le~Bolloc'h et~al.}(2016)\citenamefont{Le~Bolloc'h,
  Sinchenko, Jacques, Ortega, Lorenzo, Chahine, Lejay, and
  Monceau}}]{LeBolloch2016}
\bibinfo{author}{\bibfnamefont{D.}~\bibnamefont{Le~Bolloc'h}},
  \bibinfo{author}{\bibfnamefont{A.~A.} \bibnamefont{Sinchenko}},
  \bibinfo{author}{\bibfnamefont{V.~L.~R.} \bibnamefont{Jacques}},
  \bibinfo{author}{\bibfnamefont{L.}~\bibnamefont{Ortega}},
  \bibinfo{author}{\bibfnamefont{J.~E.} \bibnamefont{Lorenzo}},
  \bibinfo{author}{\bibfnamefont{G.~A.} \bibnamefont{Chahine}},
  \bibinfo{author}{\bibfnamefont{P.}~\bibnamefont{Lejay}}, \bibnamefont{and}
  \bibinfo{author}{\bibfnamefont{P.}~\bibnamefont{Monceau}},
  \bibinfo{journal}{Phys. Rev. B} \textbf{\bibinfo{volume}{93}},
  \bibinfo{pages}{165124} (\bibinfo{year}{2016}).

\bibitem[{\citenamefont{Overhauser}(1962)}]{overhauser1962}
\bibinfo{author}{\bibfnamefont{A.~W.} \bibnamefont{Overhauser}},
  \bibinfo{journal}{Phys. Rev.} \textbf{\bibinfo{volume}{128}},
  \bibinfo{pages}{1437} (\bibinfo{year}{1962}).

\bibitem[{\citenamefont{Young and Sokoloff}(1974)}]{Young1974}
\bibinfo{author}{\bibfnamefont{C.}~\bibnamefont{Young}} \bibnamefont{and}
  \bibinfo{author}{\bibfnamefont{J.}~\bibnamefont{Sokoloff}},
  \bibinfo{journal}{Journal of Physics F: Metal Physics}
  \textbf{\bibinfo{volume}{4}}, \bibinfo{pages}{1304} (\bibinfo{year}{1974}).

\bibitem[{\citenamefont{Tsunoda et~al.}(1974)\citenamefont{Tsunoda, Mori,
  Kunitomi, Teraoka, and Kanamori}}]{Tsunoda1974}
\bibinfo{author}{\bibfnamefont{Y.}~\bibnamefont{Tsunoda}},
  \bibinfo{author}{\bibfnamefont{M.}~\bibnamefont{Mori}},
  \bibinfo{author}{\bibfnamefont{N.}~\bibnamefont{Kunitomi}},
  \bibinfo{author}{\bibfnamefont{Y.}~\bibnamefont{Teraoka}}, \bibnamefont{and}
  \bibinfo{author}{\bibfnamefont{J.}~\bibnamefont{Kanamori}},
  \bibinfo{journal}{Solid State Communications} \textbf{\bibinfo{volume}{14}},
  \bibinfo{pages}{287 } (\bibinfo{year}{1974}).

\bibitem[{\citenamefont{Corliss et~al.}(1959)\citenamefont{Corliss, Hastings,
  and Weiss}}]{Corliss1959}
\bibinfo{author}{\bibfnamefont{L.~M.} \bibnamefont{Corliss}},
  \bibinfo{author}{\bibfnamefont{J.~M.} \bibnamefont{Hastings}},
  \bibnamefont{and} \bibinfo{author}{\bibfnamefont{R.~J.} \bibnamefont{Weiss}},
  \bibinfo{journal}{Phys. Rev. Lett.} \textbf{\bibinfo{volume}{3}},
  \bibinfo{pages}{211} (\bibinfo{year}{1959}).

\bibitem[{\citenamefont{Werner et~al.}(1967)\citenamefont{Werner, Arrott, and
  Kendrick}}]{Werner1967}
\bibinfo{author}{\bibfnamefont{S.~A.} \bibnamefont{Werner}},
  \bibinfo{author}{\bibfnamefont{A.}~\bibnamefont{Arrott}}, \bibnamefont{and}
  \bibinfo{author}{\bibfnamefont{H.}~\bibnamefont{Kendrick}},
  \bibinfo{journal}{Phys. Rev.} \textbf{\bibinfo{volume}{155}},
  \bibinfo{pages}{528} (\bibinfo{year}{1967}).

\bibitem[{\citenamefont{Cowan}(1978)}]{Cowan1978}
\bibinfo{author}{\bibfnamefont{W.}~\bibnamefont{Cowan}},
  \bibinfo{journal}{Journal of Physics F: Metal Physics}
  \textbf{\bibinfo{volume}{8}}, \bibinfo{pages}{423} (\bibinfo{year}{1978}).

\bibitem[{\citenamefont{Jacques et~al.}(2009)\citenamefont{Jacques, Bolloc'h,
  Ravy, Giles, Livet, and Wilkins}}]{Jacques2009b}
\bibinfo{author}{\bibfnamefont{V.~L.} \bibnamefont{Jacques}},
  \bibinfo{author}{\bibfnamefont{D.~L.} \bibnamefont{Bolloc'h}},
  \bibinfo{author}{\bibfnamefont{S.}~\bibnamefont{Ravy}},
  \bibinfo{author}{\bibfnamefont{C.}~\bibnamefont{Giles}},
  \bibinfo{author}{\bibfnamefont{F.}~\bibnamefont{Livet}}, \bibnamefont{and}
  \bibinfo{author}{\bibfnamefont{S.~B.} \bibnamefont{Wilkins}},
  \bibinfo{journal}{Eur. Phys. J. B} \textbf{\bibinfo{volume}{70}},
  \bibinfo{pages}{317} (\bibinfo{year}{2009}).

\bibitem[{\citenamefont{Jacques et~al.}(2014)\citenamefont{Jacques, Pinsolle,
  Ravy, Abramovici, and Le~Bolloc'h}}]{Jacques2014}
\bibinfo{author}{\bibfnamefont{V.~L.~R.} \bibnamefont{Jacques}},
  \bibinfo{author}{\bibfnamefont{E.}~\bibnamefont{Pinsolle}},
  \bibinfo{author}{\bibfnamefont{S.}~\bibnamefont{Ravy}},
  \bibinfo{author}{\bibfnamefont{G.}~\bibnamefont{Abramovici}},
  \bibnamefont{and}
  \bibinfo{author}{\bibfnamefont{D.}~\bibnamefont{Le~Bolloc'h}},
  \bibinfo{journal}{Phys. Rev. B} \textbf{\bibinfo{volume}{89}},
  \bibinfo{pages}{245127} (\bibinfo{year}{2014}).

\bibitem[{\citenamefont{Evans et~al.}(2002)\citenamefont{Evans, Isaacs, Aeppli,
  Cai, and Lai}}]{Evans2002}
\bibinfo{author}{\bibfnamefont{P.}~\bibnamefont{Evans}},
  \bibinfo{author}{\bibfnamefont{E.}~\bibnamefont{Isaacs}},
  \bibinfo{author}{\bibfnamefont{G.}~\bibnamefont{Aeppli}},
  \bibinfo{author}{\bibfnamefont{Z.}~\bibnamefont{Cai}}, \bibnamefont{and}
  \bibinfo{author}{\bibfnamefont{B.}~\bibnamefont{Lai}},
  \bibinfo{journal}{Science} \textbf{\bibinfo{volume}{295}},
  \bibinfo{pages}{1042} (\bibinfo{year}{2002}).

\bibitem[{\citenamefont{Yusupov et~al.}(2010)\citenamefont{Yusupov, Mertelj,
  Kabanov, Brazovskii, Kusar, Chu, Fisher, and Mihailovic}}]{Yusupov2010}
\bibinfo{author}{\bibfnamefont{R.}~\bibnamefont{Yusupov}},
  \bibinfo{author}{\bibfnamefont{T.}~\bibnamefont{Mertelj}},
  \bibinfo{author}{\bibfnamefont{V.~V.} \bibnamefont{Kabanov}},
  \bibinfo{author}{\bibfnamefont{S.}~\bibnamefont{Brazovskii}},
  \bibinfo{author}{\bibfnamefont{P.}~\bibnamefont{Kusar}},
  \bibinfo{author}{\bibfnamefont{J.-H.} \bibnamefont{Chu}},
  \bibinfo{author}{\bibfnamefont{I.~R.} \bibnamefont{Fisher}},
  \bibnamefont{and}
  \bibinfo{author}{\bibfnamefont{D.}~\bibnamefont{Mihailovic}},
  \bibinfo{journal}{Nat Phys} \textbf{\bibinfo{volume}{6}},
  \bibinfo{pages}{681} (\bibinfo{year}{2010}).

\bibitem[{\citenamefont{Demsar et~al.}(1999)\citenamefont{Demsar, Biljakovic,
  and Mihailovic}}]{Demsar1999}
\bibinfo{author}{\bibfnamefont{J.}~\bibnamefont{Demsar}},
  \bibinfo{author}{\bibfnamefont{K.}~\bibnamefont{Biljakovic}},
  \bibnamefont{and}
  \bibinfo{author}{\bibfnamefont{D.}~\bibnamefont{Mihailovic}},
  \bibinfo{journal}{Phys. Rev. Lett.} \textbf{\bibinfo{volume}{83}},
  \bibinfo{pages}{800} (\bibinfo{year}{1999}).

\bibitem[{\citenamefont{Tomeljak et~al.}(2009)\citenamefont{Tomeljak, Schafer,
  Stadter, Beyer, Biljakovic, and Demsar}}]{Tomeljak2009}
\bibinfo{author}{\bibfnamefont{A.}~\bibnamefont{Tomeljak}},
  \bibinfo{author}{\bibfnamefont{H.}~\bibnamefont{Schafer}},
  \bibinfo{author}{\bibfnamefont{D.}~\bibnamefont{Stadter}},
  \bibinfo{author}{\bibfnamefont{M.}~\bibnamefont{Beyer}},
  \bibinfo{author}{\bibfnamefont{K.}~\bibnamefont{Biljakovic}},
  \bibnamefont{and} \bibinfo{author}{\bibfnamefont{J.}~\bibnamefont{Demsar}},
  \bibinfo{journal}{Phys. Rev. Lett.} \textbf{\bibinfo{volume}{102}},
  \bibinfo{pages}{066404} (\bibinfo{year}{2009}).

\bibitem[{\citenamefont{Schmitt et~al.}(2008)\citenamefont{Schmitt, Kirchmann,
  Bovensiepen, Moore, Rettig, Krenz, Chu, Ru, Perfetti, Lu
  et~al.}}]{Schmitt2008}
\bibinfo{author}{\bibfnamefont{F.}~\bibnamefont{Schmitt}},
  \bibinfo{author}{\bibfnamefont{P.~S.} \bibnamefont{Kirchmann}},
  \bibinfo{author}{\bibfnamefont{U.}~\bibnamefont{Bovensiepen}},
  \bibinfo{author}{\bibfnamefont{R.~G.} \bibnamefont{Moore}},
  \bibinfo{author}{\bibfnamefont{L.}~\bibnamefont{Rettig}},
  \bibinfo{author}{\bibfnamefont{M.}~\bibnamefont{Krenz}},
  \bibinfo{author}{\bibfnamefont{J.-H.} \bibnamefont{Chu}},
  \bibinfo{author}{\bibfnamefont{N.}~\bibnamefont{Ru}},
  \bibinfo{author}{\bibfnamefont{L.}~\bibnamefont{Perfetti}},
  \bibinfo{author}{\bibfnamefont{D.~H.} \bibnamefont{Lu}},
  \bibnamefont{et~al.}, \bibinfo{journal}{Science}
  \textbf{\bibinfo{volume}{321}}, \bibinfo{pages}{1649} (\bibinfo{year}{2008}).

\bibitem[{\citenamefont{Liu et~al.}(2013)\citenamefont{Liu, Gierz, Petersen,
  Kaiser, Simoncig, Cavalieri, Cacho, Turcu, Springate, Frassetto
  et~al.}}]{Liu2013}
\bibinfo{author}{\bibfnamefont{H.~Y.} \bibnamefont{Liu}},
  \bibinfo{author}{\bibfnamefont{I.}~\bibnamefont{Gierz}},
  \bibinfo{author}{\bibfnamefont{J.~C.} \bibnamefont{Petersen}},
  \bibinfo{author}{\bibfnamefont{S.}~\bibnamefont{Kaiser}},
  \bibinfo{author}{\bibfnamefont{A.}~\bibnamefont{Simoncig}},
  \bibinfo{author}{\bibfnamefont{A.~L.} \bibnamefont{Cavalieri}},
  \bibinfo{author}{\bibfnamefont{C.}~\bibnamefont{Cacho}},
  \bibinfo{author}{\bibfnamefont{I.~C.~E.} \bibnamefont{Turcu}},
  \bibinfo{author}{\bibfnamefont{E.}~\bibnamefont{Springate}},
  \bibinfo{author}{\bibfnamefont{F.}~\bibnamefont{Frassetto}},
  \bibnamefont{et~al.}, \bibinfo{journal}{Phys. Rev. B}
  \textbf{\bibinfo{volume}{88}}, \bibinfo{pages}{045104}
  (\bibinfo{year}{2013}).

\bibitem[{\citenamefont{Eichberger et~al.}(2010)\citenamefont{Eichberger,
  Schafer, Krumova, Beyer, Demsar, Berger, Moriena, Sciaini, and
  Miller}}]{Eichberger2010}
\bibinfo{author}{\bibfnamefont{M.}~\bibnamefont{Eichberger}},
  \bibinfo{author}{\bibfnamefont{H.}~\bibnamefont{Schafer}},
  \bibinfo{author}{\bibfnamefont{M.}~\bibnamefont{Krumova}},
  \bibinfo{author}{\bibfnamefont{M.}~\bibnamefont{Beyer}},
  \bibinfo{author}{\bibfnamefont{J.}~\bibnamefont{Demsar}},
  \bibinfo{author}{\bibfnamefont{H.}~\bibnamefont{Berger}},
  \bibinfo{author}{\bibfnamefont{G.}~\bibnamefont{Moriena}},
  \bibinfo{author}{\bibfnamefont{G.}~\bibnamefont{Sciaini}}, \bibnamefont{and}
  \bibinfo{author}{\bibfnamefont{R.~J.~D.} \bibnamefont{Miller}},
  \bibinfo{journal}{Nature} \textbf{\bibinfo{volume}{468}},
  \bibinfo{pages}{799} (\bibinfo{year}{2010}).

\bibitem[{\citenamefont{Huber et~al.}(2014)\citenamefont{Huber, Mariager,
  Ferrer, Sch\"afer, Johnson, Gr\"ubel, L\"ubcke, Huber, Kubacka, Dornes
  et~al.}}]{huber2014}
\bibinfo{author}{\bibfnamefont{T.}~\bibnamefont{Huber}},
  \bibinfo{author}{\bibfnamefont{S.}~\bibnamefont{Mariager}},
  \bibinfo{author}{\bibfnamefont{A.}~\bibnamefont{Ferrer}},
  \bibinfo{author}{\bibfnamefont{H.}~\bibnamefont{Sch\"afer}},
  \bibinfo{author}{\bibfnamefont{J.}~\bibnamefont{Johnson}},
  \bibinfo{author}{\bibfnamefont{S.}~\bibnamefont{Gr\"ubel}},
  \bibinfo{author}{\bibfnamefont{A.}~\bibnamefont{L\"ubcke}},
  \bibinfo{author}{\bibfnamefont{L.}~\bibnamefont{Huber}},
  \bibinfo{author}{\bibfnamefont{T.}~\bibnamefont{Kubacka}},
  \bibinfo{author}{\bibfnamefont{C.}~\bibnamefont{Dornes}},
  \bibnamefont{et~al.}, \bibinfo{journal}{Phys. Rev. Lett.}
  \textbf{\bibinfo{volume}{113}}, \bibinfo{pages}{026401}
  (\bibinfo{year}{2014}).

\bibitem[{\citenamefont{Laulh\'e et~al.}(2015)\citenamefont{Laulh\'e, Cario,
  Corraze, Janod, Huber, Lantz, Boulfaat, Ferrer, Mariager, Johnson
  et~al.}}]{laulhe2015}
\bibinfo{author}{\bibfnamefont{C.}~\bibnamefont{Laulh\'e}},
  \bibinfo{author}{\bibfnamefont{L.}~\bibnamefont{Cario}},
  \bibinfo{author}{\bibfnamefont{B.}~\bibnamefont{Corraze}},
  \bibinfo{author}{\bibfnamefont{E.}~\bibnamefont{Janod}},
  \bibinfo{author}{\bibfnamefont{T.}~\bibnamefont{Huber}},
  \bibinfo{author}{\bibfnamefont{G.}~\bibnamefont{Lantz}},
  \bibinfo{author}{\bibfnamefont{S.}~\bibnamefont{Boulfaat}},
  \bibinfo{author}{\bibfnamefont{A.}~\bibnamefont{Ferrer}},
  \bibinfo{author}{\bibfnamefont{S.}~\bibnamefont{Mariager}},
  \bibinfo{author}{\bibfnamefont{J.}~\bibnamefont{Johnson}},
  \bibnamefont{et~al.}, \bibinfo{journal}{Physica B: Condensed Matter}
  \textbf{\bibinfo{volume}{460}}, \bibinfo{pages}{100} (\bibinfo{year}{2015}).

\bibitem[{\citenamefont{Hirori et~al.}(2003)\citenamefont{Hirori, Tachizaki,
  Matsuda, and Wright}}]{hirori2003}
\bibinfo{author}{\bibfnamefont{H.}~\bibnamefont{Hirori}},
  \bibinfo{author}{\bibfnamefont{T.}~\bibnamefont{Tachizaki}},
  \bibinfo{author}{\bibfnamefont{O.}~\bibnamefont{Matsuda}}, \bibnamefont{and}
  \bibinfo{author}{\bibfnamefont{O.~B.} \bibnamefont{Wright}},
  \bibinfo{journal}{Phys. Rev. B} \textbf{\bibinfo{volume}{68}},
  \bibinfo{pages}{113102} (\bibinfo{year}{2003}).

\bibitem[{\citenamefont{Brorson et~al.}(1990)\citenamefont{Brorson,
  Kazeroonian, Moodera, Face, Cheng, Ippen, Dresselhaus, and
  Dresselhaus}}]{Brorson1990}
\bibinfo{author}{\bibfnamefont{S.~D.} \bibnamefont{Brorson}},
  \bibinfo{author}{\bibfnamefont{A.}~\bibnamefont{Kazeroonian}},
  \bibinfo{author}{\bibfnamefont{J.~S.} \bibnamefont{Moodera}},
  \bibinfo{author}{\bibfnamefont{D.~W.} \bibnamefont{Face}},
  \bibinfo{author}{\bibfnamefont{T.~K.} \bibnamefont{Cheng}},
  \bibinfo{author}{\bibfnamefont{E.~P.} \bibnamefont{Ippen}},
  \bibinfo{author}{\bibfnamefont{M.~S.} \bibnamefont{Dresselhaus}},
  \bibnamefont{and}
  \bibinfo{author}{\bibfnamefont{G.}~\bibnamefont{Dresselhaus}},
  \bibinfo{journal}{Phys. Rev. Lett.} \textbf{\bibinfo{volume}{64}},
  \bibinfo{pages}{2172} (\bibinfo{year}{1990}).

\bibitem[{\citenamefont{Singer et~al.}(2015)\citenamefont{Singer, Marsh,
  Dietze, Uhlíř, Li, Walko, Dufresne, Srajer, Cosgriff, Evans
  et~al.}}]{singer2015}
\bibinfo{author}{\bibfnamefont{A.}~\bibnamefont{Singer}},
  \bibinfo{author}{\bibfnamefont{M.~J.} \bibnamefont{Marsh}},
  \bibinfo{author}{\bibfnamefont{S.~H.} \bibnamefont{Dietze}},
  \bibinfo{author}{\bibfnamefont{V.}~\bibnamefont{Uhlíř}},
  \bibinfo{author}{\bibfnamefont{Y.}~\bibnamefont{Li}},
  \bibinfo{author}{\bibfnamefont{D.~A.} \bibnamefont{Walko}},
  \bibinfo{author}{\bibfnamefont{E.~M.} \bibnamefont{Dufresne}},
  \bibinfo{author}{\bibfnamefont{G.}~\bibnamefont{Srajer}},
  \bibinfo{author}{\bibfnamefont{M.~P.} \bibnamefont{Cosgriff}},
  \bibinfo{author}{\bibfnamefont{P.~G.} \bibnamefont{Evans}},
  \bibnamefont{et~al.}, \bibinfo{journal}{Phys. Rev. B}
  \textbf{\bibinfo{volume}{91}}, \bibinfo{pages}{115134}
  (\bibinfo{year}{2015}).

\bibitem[{\citenamefont{Feinberg and Friedel}(1988)}]{Feinberg1988}
\bibinfo{author}{\bibfnamefont{D.}~\bibnamefont{Feinberg}} \bibnamefont{and}
  \bibinfo{author}{\bibfnamefont{J.}~\bibnamefont{Friedel}},
  \bibinfo{journal}{J. Phys. France} \textbf{\bibinfo{volume}{49}},
  \bibinfo{pages}{485} (\bibinfo{year}{1988}).

\bibitem[{\citenamefont{Isakovic et~al.}(2006)\citenamefont{Isakovic, Evans,
  Kmetko, Cicak, Cai, Lai, and Thorne}}]{Isakovic2006}
\bibinfo{author}{\bibfnamefont{A.~F.} \bibnamefont{Isakovic}},
  \bibinfo{author}{\bibfnamefont{P.~G.} \bibnamefont{Evans}},
  \bibinfo{author}{\bibfnamefont{J.}~\bibnamefont{Kmetko}},
  \bibinfo{author}{\bibfnamefont{K.}~\bibnamefont{Cicak}},
  \bibinfo{author}{\bibfnamefont{Z.}~\bibnamefont{Cai}},
  \bibinfo{author}{\bibfnamefont{B.}~\bibnamefont{Lai}}, \bibnamefont{and}
  \bibinfo{author}{\bibfnamefont{R.~E.} \bibnamefont{Thorne}},
  \bibinfo{journal}{Phys. Rev. Lett.} \textbf{\bibinfo{volume}{96}},
  \bibinfo{pages}{046401} (\bibinfo{year}{2006}).

\end{thebibliography}

\end{document}